# Review of the some specific features of the detecting of heavy recoils

**Yu.S. Tsyganov[a*], D. Ibadullayev[a, b, c], A.N. Polyakov[a], A.A. Voinov[a], M.V. Shumeiko[a], V.A. Shubin[a], V.B. Zlokazov[a], D.A. Kuznetsov[a].**

[a]*Joint Institute for Nuclear Research, 141980 Dubna, Russian Federation.*

[b]*Institute of Nuclear Physics, 050032 Almaty, Kazakhstan.*

[c]*L.N. Gumilyov Eurasian National University, 010000 Astana, Kazakhstan.*

*Corresponding author

*Email*: tyra@jinr.ru

ABSTRACT: In this paper, we present the results of the first beam tests of the detection system at the focal plane of the Dubna Gas-Filled Recoil Separator-2 (DGFRS-2), which receives beams from the DC-280 FLNR cyclotron. The high beam intensity of $^{48}$Ca$^{+10}$ heavy ions from the cyclotron enables us to obtain a number of superheavy recoils sufficient to compare both the measured and calculated spectra of superheavy recoils implanted into a silicon detector. A real-time algorithm to search for an Evaporation Residue (ER) - α correlated sequences is described in brief. It should be noted that the DGFRS-2 spectrometer operates in conjunction with the 48×128 strip DSSD (Double-sided Silicon Strip Detector; 48×226 mm$^2$) detector and a low-pressure pentane-filled gaseous detector (1.2 Torr; 80×230 mm$^2$). A block-diagram of the spectrometer and the event format are also presented. Special attention is paid to the response of a low-pressure pentane-filled ΔE multiwire proportional chamber for recoils of Fl, synthesized in the $^{242}$Pu+$^{48}$Ca→$^{287}$Fl +3n complete fusion nuclear reaction. Some actual parameters of the detection system have also been extracted from $^{nat}$Yb + $^{48}$Ca, $^{232}$Th + $^{48}$Ca, $^{243}$Am + $^{48}$Ca, $^{238}$U + $^{48}$Ca reactions. The effect of neighbor strip charge sharing for the ohmic side of the DSSD detector is also under consideration.

## Introduction

The synthesis of new elements with atomic numbers 113 to 118 was achieved through the use of gas-filled electromagnetic separators and corresponding detection systems in experiments with $^{48}$Ca ions [1-4]. Improved detection systems allowed for the isolation of rare alpha and spontaneous fission decays of super heavy nuclei (SHN) from background events in reactions such as $^{48}$Ca + actinide target → SHN + xn performed at the U-400 cyclotron, FLNR JINR [4-5]. The cross sections for these reactions range from 0.1 to 10 picobarns. However, with the use of $^{50}$Ti and $^{54}$Cr ion beams in future experiments, a significant decrease in SHN production cross-sections is expected, which raises the requirements for the properties of the separator and detection system. As a result, the task of studying the features of detecting rare decay events of superheavy nuclei becomes increasingly important. Additionally, the method of active correlations [4, 6-9], which is used to suppress background, is especially crucial when using intense beams of heavy ions (up to 5-10 pμA).

## 1. Separator DGFRS-2

In 2020, the new gas-filled recoil separator DGFRS-2 was put into operation [10-11]. The DGFRS-II is intended for separating heavy nuclei — the products of fusion–evaporation reactions — from a beam of heavy ions bombarding a target and other background particles. Unlike its predecessor, the DGFRS-1 separator [1], it has a more advanced scheme for separating the products of nuclear reactions (see Fig. 1). The desired intensity of heavy ion beams in experiments can reach ~ 10 pµA, instead of about ~ 1 pµA in experiments on the DGFRS-1 separator [12-15].

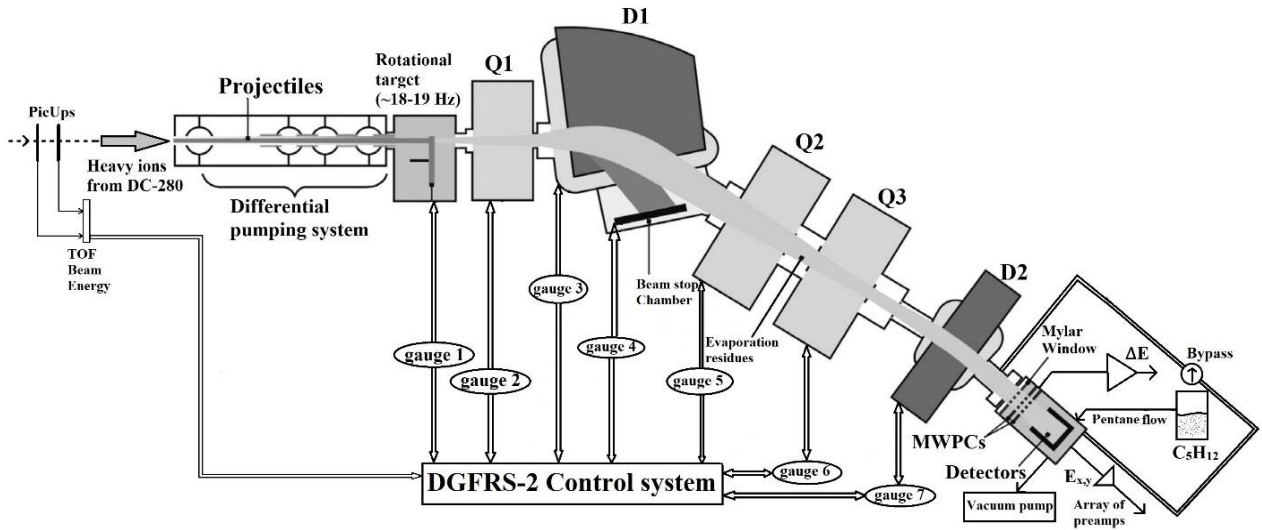

**Figure. 1.** Installation of the DGFRS-2.

The installation of the DGFRS-2 features differential pumping, magnetic dipoles, lenses, and a detection module. The pentane subsystem is equipped with a dry vacuum pump (TriScrol Agilent, PTS 600). The Mylar window separates pentane in the detection system from hydrogen in the separator. A bypass system is in place to prevent damage to the Mylar window in the event of high hydrogen pressure (>1 Torr), high pentane pressure (> 8 Torr), or low pentane pressure (< 1 Torr).

The main feature of this installation is the high efficiency of collecting synthesized superheavy nuclei, which is estimated to be about 60% for targets with a thickness of 0.5 mg/cm$^2$. The new separator provides higher background suppression compared to the old one by the factor of about 200 (see Fig. 10 in [10]). As seen in Fig.1 the DGFRS-II separator consists of 5 magnetic elements: 2 dipole magnets (D1 and D2) designed to separate the products of the complete fusion reaction from background particles, and 3 quadrupole lenses (Q1-Q3) that focus the reaction products in the focal plane of the separator, where the detecting system is located.

## 2. Detection of implanted recoil nuclei

The DGFRS-2 detection system, shown in Fig. 2a, registers the signals of implanted evaporation residues (ER) and their subsequent decay products. A 48x220 mm$^2$ Double-sided Silicon Strip Detector (DSSD) with 48(front 1 mm wide) × 128(back 2 mm wide) strips acts as the implantation detector. Eight side detectors each with eight strips (64 strips 15 mm wide) record particles that escape from the implantation DSSD in the upstream direction. Further upstream is placed a Multi-Wire Proportional Counter (MWPC), which provides a signal with which ERs, detected in the DSSD, can be distinguished from their subsequent decay. It also allows for a ΔE measurement that can be used to clean SHN from the background, which will be discussed later. Some attributes of MWPC: Cathode 20 µm W(Au) wire, step 1 mm; Anode 10 µm W(Au) wire, step 2 mm. In addition, two detectors, placed behind the DSSD, veto background events that punch through the DSSD. The active correlations method [4, 6-9], used for radical background suppression, requires that the beam be stopped immediately (30 µs) after a candidate ER-alpha correlation is found in real-time. Therefore, the subsequent decay of the potential SHN chain occurs during a beam-off period. It is crucial that the ER signal be well identified, as this opens the time window in which potential decay chains are searched. The simple yet effective idea behind this method is that the PC-based Builder C++ program searches in real-time mode for energy-time-position recoil-alpha correlations (ER-α), using two matrix representations of the DSSD detector, one for ER matrix and one for α-matrix. Upon detection of an "alpha particle" signal, a comparison is made with the appropriate "ER-matrix" element. If the elapsed time difference between the "recoil" and the "alpha particle" is within the preset time value, the system turns on the cyclotron beam chopper, which deflects the heavy ion beam in the injection line of the DC-280 FLNR cyclotron for a predefined time interval (usually 0.5-2 min). The next step of the computer C++ code is to detect of the subsequent α-particle signals during the beam-off interval. If such a decay takes place in the same position (X, Y), the duration of the beam-off interval is prolonged by a factor 3-10. The dead time of the system, associated with interrupting the beam, is about ~100 µs, including linear growth chopper operation delay (~10µs) and estimated heavy ion orbit life-time (~60 µs).

In addition, the time interval between ER and the first α-decay is used to estimate the half-life of the nuclide under study.

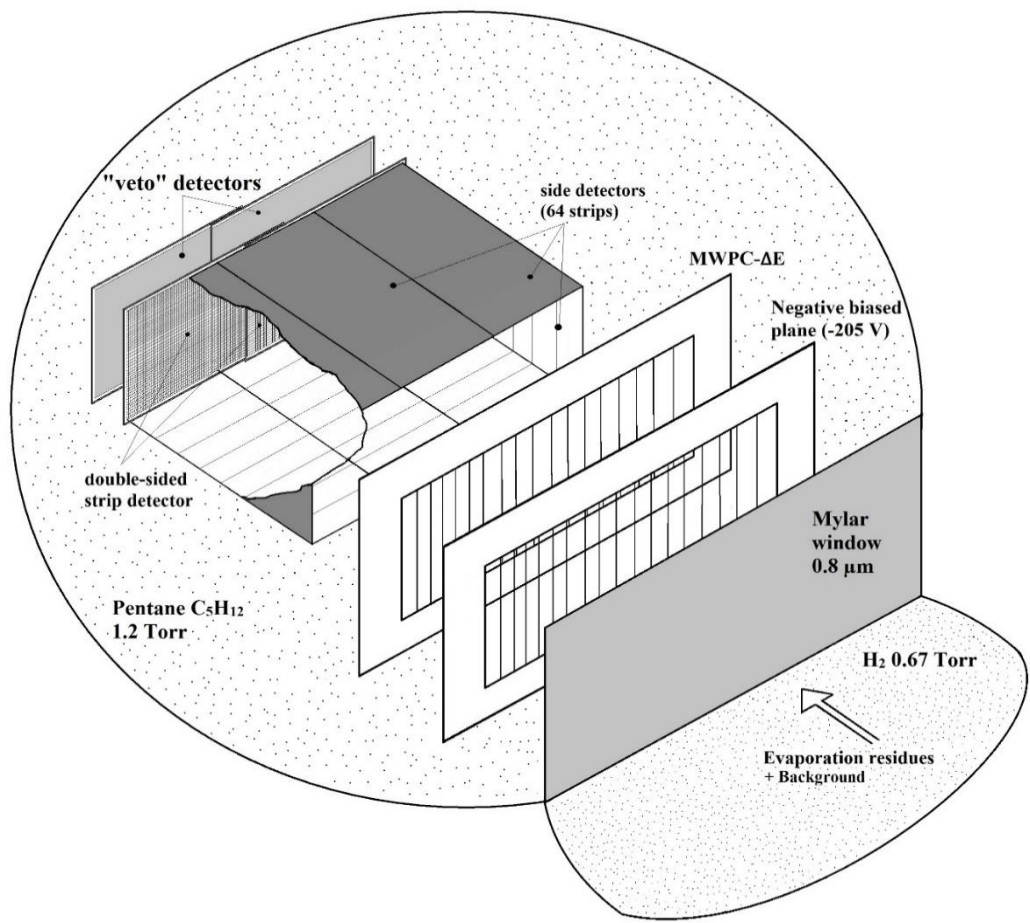

**Figure. 2a.** Schematic of the DGFRS-2 detection system module.

In order to provide the maximum purity of pentane vapor in the working volume of the ΔE chamber, we set a relatively high flow rate of about 2 liters/week. The block diagram of the pentane renewal process is shown in Fig. 2b. With this system, it is possible to preset both the pressure and gas flow value at the start of each experiment using a Pfeiffer RVC-300 PID controller, which is remotely operated from a PC. The liquid pentane vessel has a volume of 6 liters. These measures are taken to minimize the impact of hydrogen diffusing through the ultra-thin Mylar window and entering the ΔE chamber volume through the diffusion process.

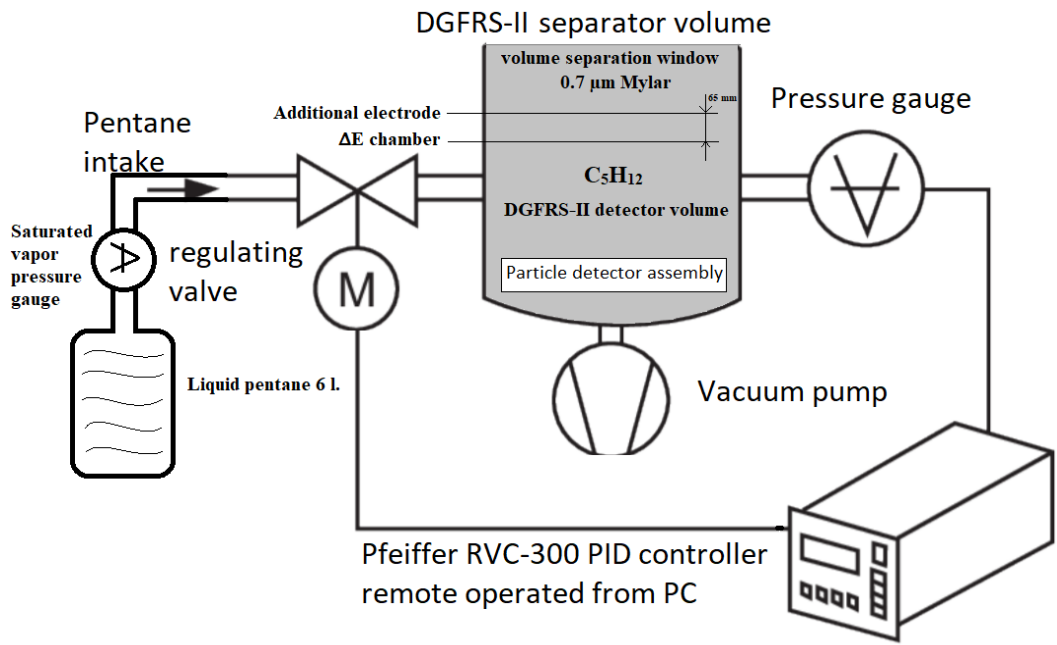

**Figure. 2b**. Gas flow subsystem of ΔE proportional chamber.

In Fig. 2c, a block-diagram for one electronics channel featuring a CAMAC crate with ADP-16 units is shown [8, 16]. Data taking process is triggered by a non-zero bit in the status register 1M unit, which corresponds to any **ADP-16 "L" TTL** signal for seven modules, three of which correspond to 48 strips of the DSSD detector, and the seven others to 64 side detectors. Gating of the Pa3n unit for reading **ΔE** signals is performed only when a fast signal from any ADP for the DSSD detector is present. The duration of the **"GATE"** signal is 6 μs. The detection system has a 4 μs local dead time due to the 8 cells ADP's FIFO (12 cells for GNS-M01), while the regular dead time is around ~30 μs. The typical resolution of the vertical strips for α-particle signals is of around 35-40 keV (*FWHM*). The initial threshold values (in channels) are written from a pre-setting text file, and the setting procedure is performed using the CAMAC function *N*A(3)F(16)* for ADP's of the DSSD horizontal strips, while for the vertical strip ADP's these values are about ~10 % greater than the ones written from the file by default. No separate timing modules are used, and all times with second and microsecond accuracy were obtained from the Windows 10 system.

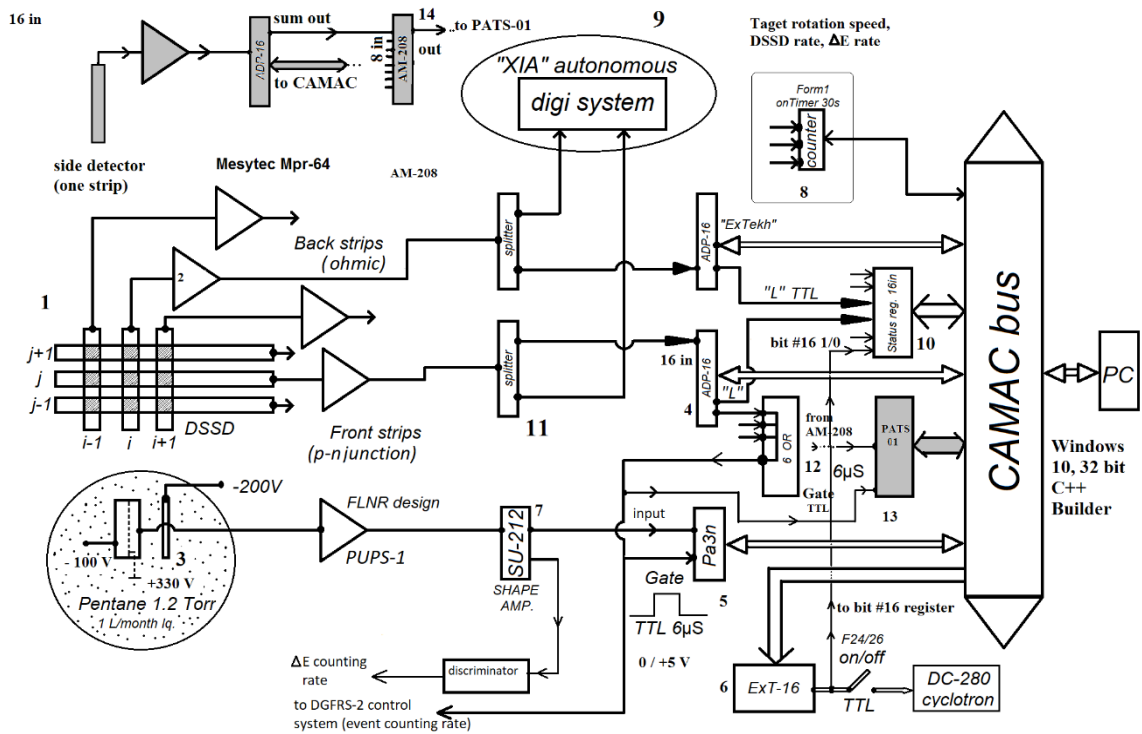

**Figure. 2c.** Block-diagram of the detection system (one electronic channel).

Block's functions of the spectrometer are listed in the Table 1.

**Table 1.** Main electronic units of the analog spectrometer (one crate system).

| Number on Figure 2c. | Unit name | Function (in brief) |
|---|---|---|
| 1 | Double Side Silicon Strip Detector | To measure ER, α, SF and other signals. |
| 2 | Mesytec Mpr-64 | Charge sensitive preamplifier. |
| 3 | ΔE low pressure pentane filled proportional chamber | To measure signal from charged particle coming from cyclotron DC-280. |
| 4 | Analog digital processor ADP-16 1M | 16-in, two scales shaping amplifier analog multiplexer-analog to digital converter: 13 bit first scale (<27 MeV), 12 bit second scale (< 270 MeV). Basic component – CPLD "ALTERA" (INTEL) Base unit of the detection system (totally -16 modules are placed in crate). Specific feature: make one microsecond synchronization between different modules using N*A (3)*F (0) function [17]. Front-back coincidences are generated online with a time interval of approximately 2.5 μs. Regular functions of ADP-16 are: N*A(0)*F(2) –readout of "α" scale (up to ~25 MeV) and |

| | | reset module; |
| --- | --- | --- |
| | | N*A(1)*F(2) – readout of "fission" scale (up to 250 MeV); |
| | | N*A(3)*F(16) – write channel of threshold value; |
| | | N(23)*A(3)*F(10) – circular reset of all ADP-16 modules in crate; |
| | | N*A(0)*F(8) – test LAM. |
| 5 | Pa3n | 3-in 1M unit to measure ΔE signals. 12 bit/channel. |
| 6 | Ext-16 | 2M unit to create "STOP" TTL signal to stop irradiation process. |
| 7 | SU-212 2M | Shaping amplifier (FLNR, JINR design) to measure ΔE signals. |
| 8 | 16 bit counter | 1M unit to measure target rotation speed, event rate of DSSD. |
| 9 | XIA digital system | XIA Corporation digital spectrometer (autonomous) (Outside the scope of this paper). |
| 10 | Status 16 bit register | Start read-out process if are there any non-zero signals of "L" any ADP of front strips (48 strips). |
| 11 | Splitter unit 3M | 32-in unit to split signals from preamplifiers to digital and analog system (FLNR, JINR design). |
| 12 | 1M unit 6OR | 6-in logical TTL input signals to provide trigger TTL signal for gating of Pa3n unit. |
| 13 | 1M unit PATS01 | 12 bit ADC to measure summary signal from all side detectors (FLNR, JINR design). |
| 14 | 1M unit AM-208 | 8- in analog multiplexer (FLNR, JINR design). |
| 15 | GNS-M-01 | The same as ADP-16, but 12 FIFO cells except for 8. It will put into operation in 2023 |

Acquisition program is written in C++ Builder [18, 19]. Format of the event is specified by fourteen 16 bit words presented in the Table 2.

Table 2. Content of the registered event (256 bit).

| W1 | W2 | W3 | W4 | W5 | W6 | W7 | W8 | W9 | W10 | W11 | W12 | W13 | W14 |
| --- | --- | --- | --- | --- | --- | --- | --- | --- | --- | --- | --- | --- | --- |
| ID | alp | fis | tsinc | sreg | alp2 | dE | fisb1 | fisb2 | cod1 | cod2 | sec | mcs | codf |

Here:

ID – type of event. 1,2,3- front focal strips, 4,5,6,7 – side detectors;

alp – amplitude of α-scale, 13 bit;

fis – 16 bit contained both codes of amplitude and strip number (amplitude 12 bit);

tsinc – time in microseconds synchronized by photodiode/light diode pair with target rotation wheel 16 bit ( usually, full cycle is about ∼ 40 ms);

sreg – 16 bit status register for one bit information. For example, bit #1 =1 means ID=1 (first ADP-16 operates );

alp2 – additional alpha particle 12 bit signal code, in the case of composite event, that is – two components are detected. One component is detected by the focal plane detector, another one, by side detector;

dE –ΔE signal, 8 bit;

fisb1 – 16 bit fission scale for back strips (8 ADP-16);

fisb2 - 16 bit fission scale for back strips in the case of sharing between two neighbor strips signal;

cod1 - 13 bit code of amplitude of α-particle signal for back strip;

cod2 – the same, if neighbor strip signal amplitude is nonzero;

sec  - Windows time in seconds starting from 00:00:00;

mcs – elapsed Windows time in microseconds, starting from file opening;

codf – fission scale code for back strips.

The ER loses noticeable energy both before implantation, passing through the working media of the separator, and due to the significant magnitude of the pulse height defect (PHD). In [20-24], the calculation of the recorded amplitudes of heavy nuclei implanted in a silicon detector is presented.

The spectrum of heavy recoils from [20] is calculated according to the scheme:

$$E_{REG} = E_0 - \Delta E_{target} - \Delta E_{H_2} - \Delta E_{Mylar} - \Delta E_{Penrane} - PHD + \Delta_{corr},$$

Here, $E_0$ is the value of the recoil energy at the middle of the target, $\Delta E_{target}$ – energy loss in the middle of the target, $\Delta E_{H_2}$ energy loss in the Hydrogen, $\Delta E_{Mylar}$ – energy loss in Mylar window, $\Delta E_{Penrane}$ – energy loss in Pentane (Fig. 3a, b), PHD-pulse height defect in silicon detector, $\Delta_{corr}$ is a small empirical correction, according to [23]. $\Delta_{corr}$ can be represented as a formula:

$$\Delta_{CORR}= 0.28 \times E_{in} - 0.006 \times E_{in}^2 - 1.08,$$

where $E_{in}$ is the recoil energy before implantation, in MeV.

The PHD value is calculated by Wilkins formula [25], whereas fluctuations due to nuclear collisions in silicon detectors are calculated according to Haines and Whitehead [26], and broadening due to neutron evaporation process is calculated according to Dahlinger et. al. [27].The energy loss in hydrogen is scaled with the current charge state $q^2$ and modeled using Chi$^2$ regularity for the charge distribution [28, 29].

Fig. 3b shows both the calculated and measured spectra of the recorded energy for $^{287}$Fl nuclei from the complete fusion reaction $^{242}$Pu($^{48}$Ca,3n)$^{287}$Fl [30, 31]. This comparing shows a satisfactory agreement between measured and calculated spectra.

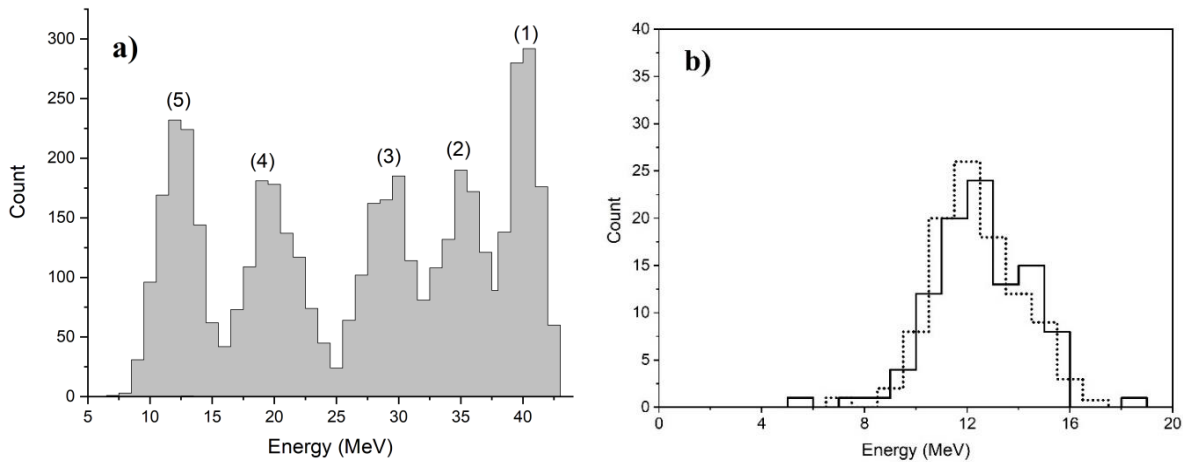

**Figure. 3.(a)** Typical spectra evolution for Fl recoil nuclei. (<$E_5$> = 11.84±1.63, <$E_4$> = 19.43±2.29, <$E_3$> = 28.72±2.27, <$E_2$> = 34.68±2.27, <$E_1$> = 39.62±1.26). **Figure. 3. (b)** Calculated (dot) and measured spectra of the $^{287}$Fl recoil at the DGFRS-2. Spectra from 1 to 5 is the spectra from target to registered energy in Si detector.

It is crucial to note that when registering the flight or implantation of the recoil nucleus, two coinciding signals are involved. The digitizer of the ΔE signal is triggered by a signal from one of the front strips of the DSSD. Maintaining a stable pressure in the low-pressure gas chamber (1.2-1.3 Torr of pentane vapor) is crucial because the chamber operates in a limited proportionality mode. Fig. 4 displays the response of the ΔE camera when registering Fl nuclei, along with a schematic representation of the spectrum of background products and target-like nuclei.

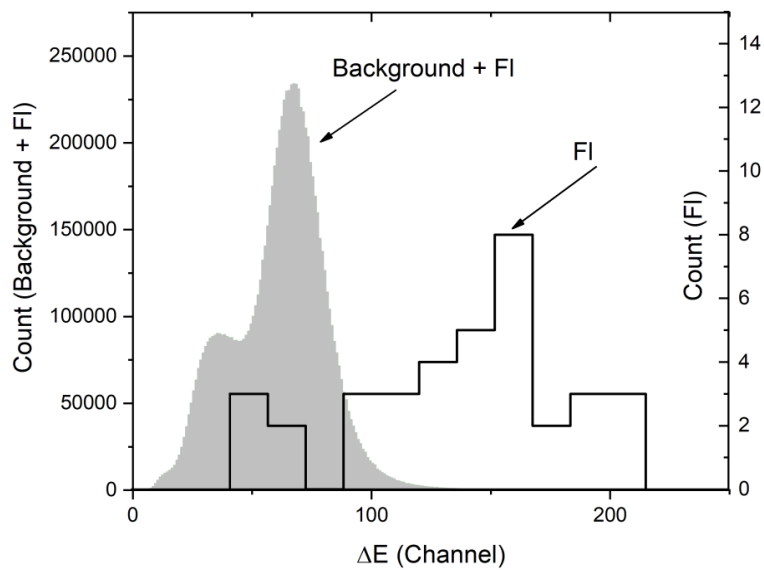

**Figure. 4.** ΔE response of the proportional low-pressure chamber. The recoil nucleus is $^{287}$Fl (grey spectrum). On the left (white spectrum) - background products plus Fl.

## 3. On the neighbor strip sharing signals

One of the specific moments in the detection of superheavy nuclei is that more than one of the 48 front strip signals may be triggered during recording. Such events are written to the data file, but are ignored in the real-time subroutine for searching for ER-α correlations that generate a short-term stop of the beam. This is done due to the negligible size of the effect, as shown in Table 3). This effect is reduced by the presence of a guard ring near each strip and a relatively low electric field, since the depletion factor of the detector (~300 μm, n-Si wafer) is not very high (45 V/ 30V=1.5). Here, 45 volts is the operating voltage, 30 volts is the depletion voltage. On the other hand, the effect of charge division on the ohmic (rear 128 strips), is relatively greater. Note, for the back side strips, with a pitch of 2 mm, the percentage of double signals for ~ 700 KeV threshold for $^{nat}Yb+^{48}Ca \rightarrow Th^*$ reaction was equal to 5.4% for 64838 events in the range 7160 to 7360 KeV.

We also note that in [22] it was shown for the alpha decay of $^{217}Th$ that the sum $E_1+E_2 = E_0$, where $E_0$, is close to the decay energy of the corresponding isotope, although with a small ~29 keV deficit (see Fig. 5).

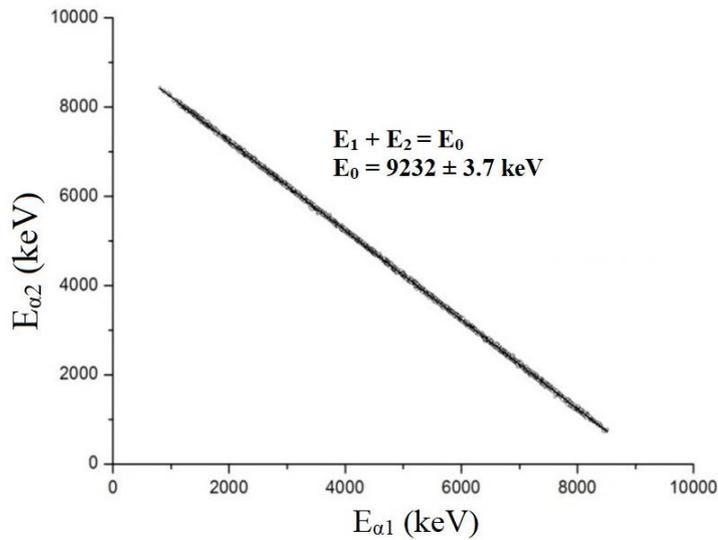

**Figure. 5.** Dependence of the energy of the adjacent strip from the side of the ohmic for the $^{217}Th$ isotope. Reaction $^{nat}Yb+^{48}Ca \rightarrow ^{217}Th +3n$ (totally 44548 events, double alpha 3545 events). $E_0 = 9232 \pm 3.7$ keV. DSSD BB-17 (Micron Scd., pitch – 1 mm (2019 year)

It is important to consider events with multiple signals from the 48 front strips during the search for ER-α correlations in real time. In this case, the value of the recoil matrix element (as described in Appendix 1 of [22]) and the elapsed time of the current event are recorded in two matrix elements: i, j and i, j+1, where i ranges from 1 to 48, and j ranges from 1 to 127.

It should be noted that the main 16 input ADP-16 modules and the GNS-01M (as listed in Table 1) operate using the principle of analog multiplexing. This means that if two or more signals are received within 300 ns, both signals are blocked. To accommodate this, the 48 front strip numbers are connected to the inputs of the ADP-16 modules as follows: 1, 2, 3, 4, 5, 6...48 connect to 1, 17, 35, 2, 18, 36... input signals.

The rear strips follow a similar pattern, but with some differences. Specifically, there is a pairwise alternation of strip numbers: 1, 2, 3, 4, 5, 6...128 connect to 1, 17, 2, 18, 3, 19... During signal processing, reverse decoding occurs.

## 4. On the issue of parameter stability

The concept of stability, although determining the success of the experiment, is a more general concept. In each particular experiment, the specific details of this general concept are of an individual nature. As it seems to us, in our particular case, first of all it is:

1) a stability of calibration parameters during several months of irradiation;

2) a stability of the vapor pressure of pentane and, as a consequence, the entire measuring channel ΔE (otherwise amplitude of ΔE signal is very unstable);

3) a stability of the parameter of the efficiency of recording the flight of recoil nuclei with a low-pressure chamber.

Note that, when using the active correlations method, a linear calibration is performed for each strip of the DSSD detector in the form $e = a_i N_{ch} + b_i$, where $a_i, b_i$ are calibration constants, $i$ is the strip number, $e$ is the measured energy, and $N_{ch}$ is the channel number (0..8191). To avoid problems caused by detector dead-layers and external alpha source coverings, the calibration coefficients are obtained from an in-beam test reaction, typically $^{nat}Yb(^{48}Ca, 3n)^{217}Th$, whereby the ER are implanted into the DSSD. The usual recalibration cycle lasts several weeks, typically 2-3. The stability of the gain constant, $a_i$, is shown in Figs. 6a and b, where comparisons are made between the values for the 48 horizontal strips at the beginning and end of a five-month-long irradiation period. For instance, this implies that for an energy of the order of 10 000 keV, the deviation will be approximately $\delta_e = 0.08/100 * 10000 = 8$ keV. This is quite satisfactory for the energy calibration, but it provides excellent stability for the active correlations method.

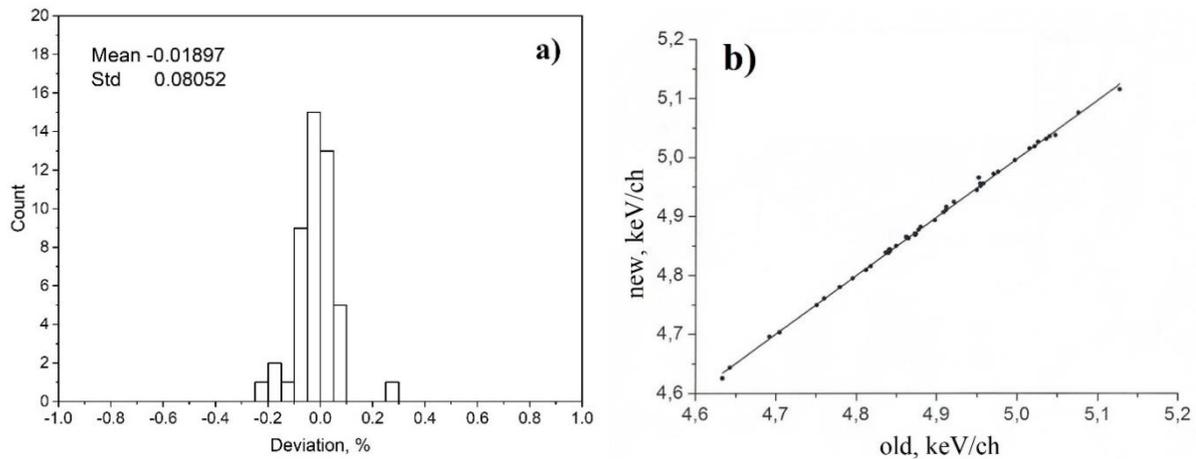

**Figure. 6. (a)** The difference in the magnitude of the calibration values of the slope of the front strips of the DSSD detector. The time distance is about 5 months. **(b)** Absolute values of slope coefficients (keV/channel). The bit depth of the ADC is 13 bits. (Correlation coefficient equal to 0.999)

As for the pentane pressure parameter (point 2), Figure 7 shows this value, which is stable for a long time during an experiment which was done on September 2022.

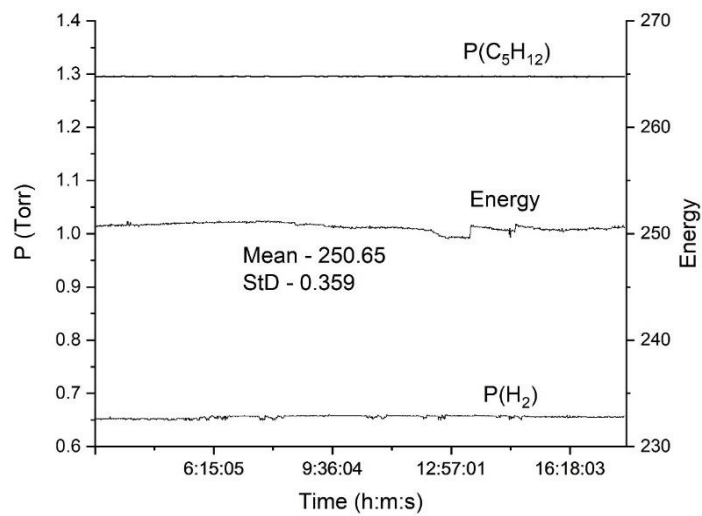

**Figure. 7.** The dependence of the pentane and hydrogen pressure value against time (approximately 15 hours. Standard deviation - 0.00045 Torr). Energy is given to show how stable it was.

The third point is slightly more complicated. Fig. 8a demonstrates that the application of a negative voltage of -205 Volts to all planes of the "start" chamber increases the amplitude of electron drift into the chamber. In reference [32], the working hypothesis is that the positive ion volume charge is responsible for the efficiency parameter dropping when the ΔE chamber count rate exceeds $2\text{-}3\cdot10^3$ s$^{-1}$, as shown in Fig. 8b. The negative voltage applied to the "start" chamber helps to remove ions from the vicinity of the working "stop" chamber and counteracts this drop in efficiency.

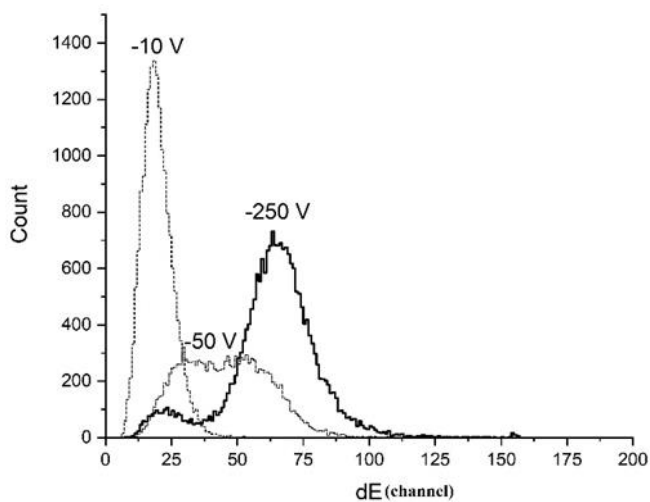

**Figure. 8a.** Change in the amplitude of the ΔE camera signal, when the drain voltage of the positive charge changes.

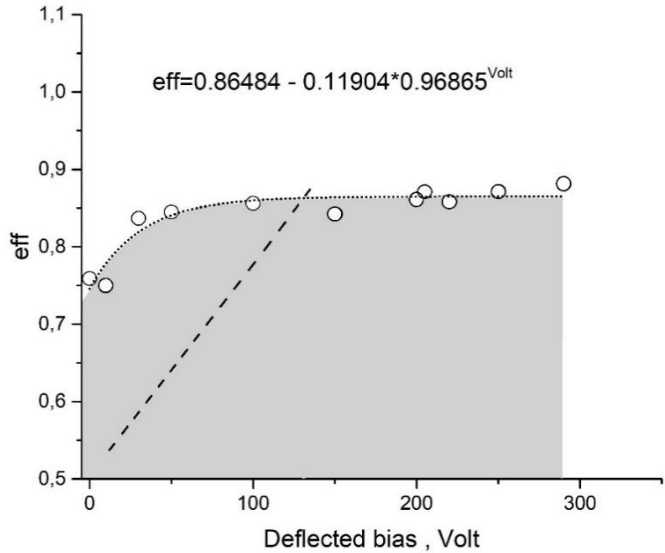

**Figure. 8b.** Dependence of the measured value of the recording efficiency on the magnitude of the deflecting voltage (reaction $^{232}$Th+$^{48}$Ca→Ds*; f = 500…1300 s$^{-1}$). Dash line indicates the trend for t ≥ $10^4$ s$^{-1}$.

Fig. 8c shows the reduced (true) value of the efficiency parameter, taking into account the fact that in the reaction of $^{232}$Th+$^{48}$Ca in the range of 5-9.5 MeV, α-decays to lead impurities predominate. This value was approximately 98.3%. At the same time, the primary, measured value (the monitoring time period of event *OnTimer* Builder C++ 5 min$^{-1}$ [19]) is defined as: $eff = \frac{N_2}{N_1}$, where $N_2$ is the total number of samples in the range of 2-4.8 MeV at a given time, and $N_1$ is a similar value, but under the condition ΔE > 0. In both cases, an additional condition is the absence of a non–zero amplitude in the side detectors (see Fig. 2) and any non-zero signal in the "VETO" detector. The choice of the energy interval is due to the desire to minimize the contribution of α-decays that perturb the correctness of the estimate.

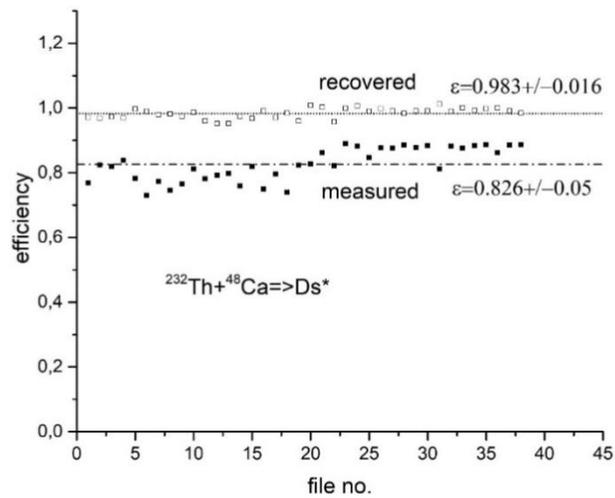

**Figure. 8c.** Measured efficiency value and restored efficiency value. Reaction $^{232}$Th+$^{48}$Ca→Ds*. The abscissa axis is the file number (arbitrary units).

The method for restoring the measured value of the ΔE camera registration efficiency is as follows: After determining the values of $N_1$ and $N_2$ for a given data file, the position and content of the alpha decay peaks for 48 front strips are interactively found using the VMRIA [33, 34] program. An example of the result of such an interactive analysis is shown in Fig. 9, which typically takes about one minute. Based on a similar analysis for an offline file (no beam, only α-decays), we determined the ratio between the number of escaping α particles in the 2-4.9 MeV range and the sum of counts in peaks. Using this information, we can determine the number of escaping α particles, $N_3$, in the working file. The corrected efficiency value is the calculated as:

$$Eff_{CORR} = \frac{N_2}{N_1 - N_3}$$

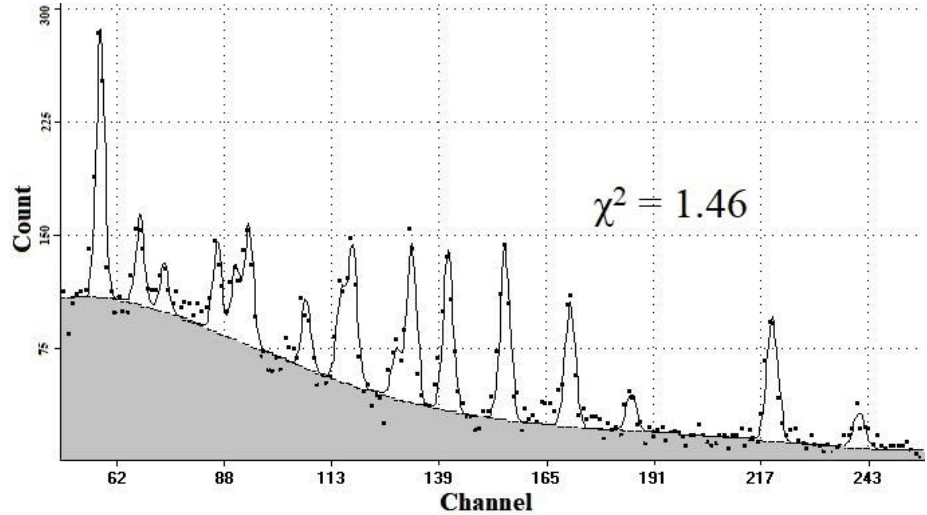

**Figure. 9.** File processing (example) with VMRIA code [33, 34]. Background marked by grey. Reaction - $^{232}$Th + $^{48}$Ca.

Positions of peaks $CP_i$ in Fig. 9 are:

{56.19±0.08, 65.80±0.21, 71.67±0.36, 84.45±0.20, 88.81±0.27, 92.02±0.17, 105.84±0.23, 114.31±0.23, 117.13±0.16, 127.69±0.30, 131.31±0.11, 140.10±0.10, 153.80±0.09, 169.51±0.11, 184.28±0.27, 218.29±0.10, 239.31±0.25}. Peak energy is equal to: $E_i = 4000.0 + 20.0 \ast CP_i$.

Results of other tests are listed in the Table 3 and show approximately the same numerical value of the efficiency monitoring parameter.

**Table. 3. Monitoring efficiency values of ΔE chamber** (*four points for each column*)

| Reaction | $^{nat}$Yb+$^{48}$Ca | $^{207}$Pb+$^{48}$Ca | $^{238}$U+$^{48}$Ca | $^{242}$Pu+$^{48}$Ca | $^{232}$Th+$^{48}$Ca |
|---|---|---|---|---|---|
| Measured efficiency | 0.98 | 0.97 | 0.98 | 0.97 | 0.83 |
| Recovered efficiency | - | 1.00 | 0.99 | - | 0.98 |
| Energy interval for efficiency test, MeV | 13-25 | 2-4.8 | 2-4.8 | 2-4.8 | 2-4.8 |

In order to provide a hypothetical explanation of the effect, refer to Fig. 10. It is clear that positively charged ions move in the direction indicated by the arrow in the field F when the potential |V| is greater than |V0|, and both potentials are negative

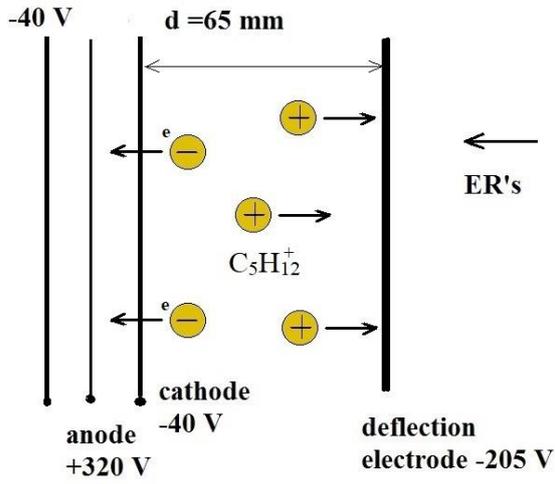

**Figure. 10.** Schematics of ΔE low pressure pentane filled chamber with deflection electrode.

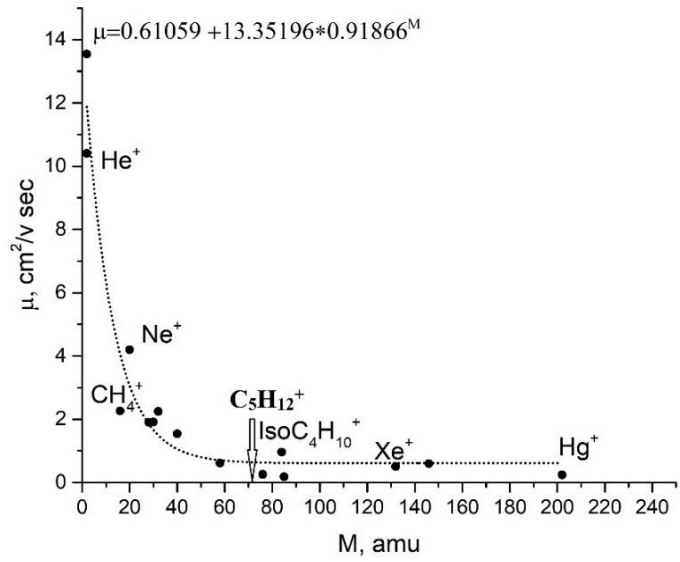

**Figure. 11.** Mobility of various ions based on [35]. Fitting of exponential function is shown.

Respectively:

$$F \approx \frac{V - V_0}{d} \to v = \mu F \to v = \frac{\mu(V - V_0)}{d} \to$$

In addition, the average charge movement time:

$$\to \tau = \frac{d}{2v} = \frac{d^2}{2\mu(V - V_0)};$$

Here, $v$ is the drift velocity of pentane ions.

On the other hand:

$$\tau = f^{-1},$$

Where $f$ is the counting rate of the ΔE chamber. Thus:

$$\frac{1}{f} = \frac{d^2}{2\mu(V-V_0)} \rightarrow V - V_0 = \frac{fd^2}{2\mu} \rightarrow \cdots \rightarrow V_{CRITICAL} = V_0 + \frac{fd^2}{2\mu}$$

Considering that under normal conditions (Fig. 11) the mobility of $C_5H_{12}^+$ ions in low pressure pentane gas $\mu_0 = 0.64\ cm^2/V \cdot s$, and, accordingly, at P=1.3 Torr,

$$\mu = \mu_0 \cdot \frac{760}{1.3} = 374\ cm^2/V \cdot s\ \text{(see [36])}.$$

- ➔ We have $V_0 = -40\ Volts$,
- ➔ Therefore $V = |V_{crit}| = 40 + \frac{6.5^2}{2 \cdot 374}f = 40 + 0.0564f$,
- ➔ $V_{crit} = V_0 + 0.0564f$.

In addition, for example:
$$f = 3 \cdot 10^3 s^{-1} \rightarrow V_{crit} = 40 + 0.0564 \cdot 3 \cdot 1000 = 40 + 169.4 = 197\ Volts$$
Of course, this estimate is approximate.

## 5. On the initial conditions for applying the active correlations technique in future experiments at the DGFRS-2 setup.

In order to detect sequences like ER-α or even ER-α-α for unknown nuclei in real-time, the researcher should use estimates, either theoretical or empirical in nature, for the lifetime of the nuclei under study, as well as for the daughter nuclei implanted into the silicon detector. In Ref.'s [16, 37], four parameters of a formula based on experimental data from Ref. [5] are presented. In [16], the results from this paper are compared with the well-known results of Royer's formula with six parameters [38]. On the other hand, during the preparation of both the manuscript and the paper [16], a very successful experiment of $^{242}Pu+^{48}Ca \rightarrow ^{287,286}Fl+3,4n$ was performed at the DGFRS-2 setup of FLNR (JINR) [30].

In principle, similar to [16], the parameter "d" was determined using the formula $\text{llog}\left(T_{\frac{1}{2}}^{measured}\right) = (aZ+b)Q_\alpha^{-\frac{1}{2}} + cZ + d$, where a=1.78722, b=-21.398, and c=-0.25488. The results are listed in Table 4 and presented in Figure 12. In this work, the parameter "d" was chosen as a free parameter, as it can be considered a more conservative parameter without a direct connection to the values of Q and Z. The experimental points, as well as four points calculated using Royer's formula, are also plotted on the graphs. The mean <d> value was found to be equal to -28.3180±0.3241. It is worth noting that this result is close to the one obtained in [16], which is d=-28.0928.

Table 4. *α*-decay parameters for six superheavy nuclei

| Isotope | $^{287}$Fl | $^{286}$Fl | $^{283}$Cn | $^{279}$Ds | $^{275}$Hs | $^{271}$Sg |
|---|---|---|---|---|---|---|
| $Q_\alpha$, MeV | 10.157 | 10.335 | 9.667 | 9.827 | 9.461 | 8.629 |
| $T_{1/2}$, s | 0.360 | 0.19 | 4.0 | 1.4 | 0.9 | 43 |
| d | -28.60256 | -28.3852 | -28.349 | -27.7 | -28.315 | -28.5565 |

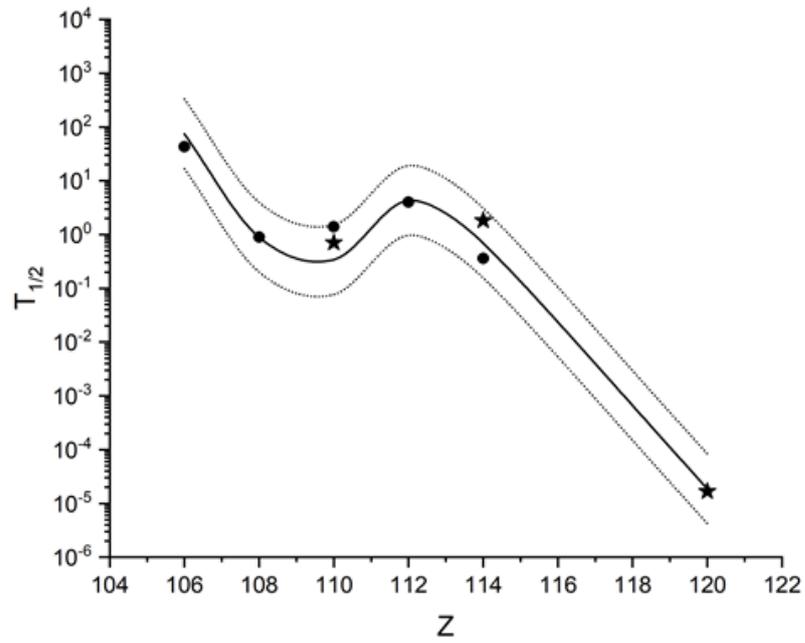

Fig. 12. Calculated $T_{1/2}$ dependence against Z. Experimental data are shown by closed circles. Royer's formula calculations are shown by stars. Dotted lines are corresponded to (±) two standard deviations.

**Supplement 1. On the realistic electrical field**

In the calculations shown before, for the sake of estimation, we have considered the electric field value in the center of the gap between the MWPC and the deflecting plane to correspond to that of an ideal capacitor. However, a small correction is presented below to account for the finite size of the MWPC (See Fig 13 a, b) [39-41].

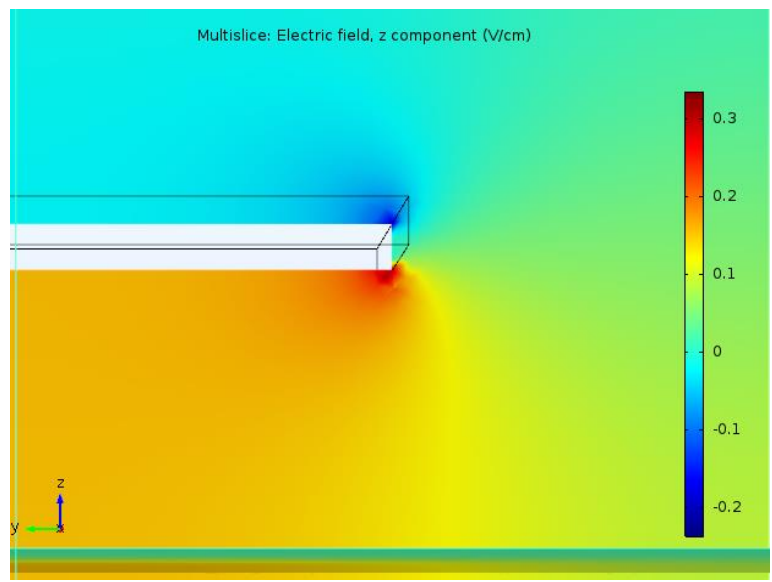

Fig. 13 a) Electrical field value distribution schematically (relative units, visually)

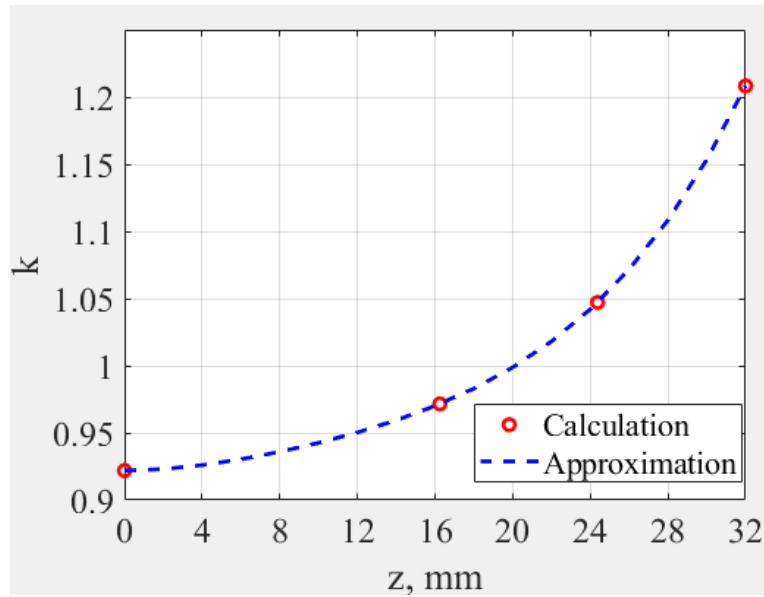

Fig. 13b) Correction coefficient for real geometry

So, in the above presented formulae, field F one should replace by $F \rightarrow k(0) \cdot F$, where k=0.92

**Summary**

With the increase in heavy ion beam intensities up to around 10 pµA (*or even more!*) m in experiments on synthesizing new elements at the DGFRS-2 facility and the DC-280 cyclotron, a number of phenomena arise that require separate study. One of them is a certain drop in the efficiency of detecting recoil nuclei with an increase in the loading parameters of the low-pressure gaseous detector ΔE. The influence of this effect is compensated by an additional deflecting voltage of an electrode located in a certain vicinity of the proportional chamber. A simple model is proposed to explain the process involving the movement of positive ions. A critical aspect of the detection process is the high stability of the electronic circuits, which allows for recalibration a limited number of times during months of experimentation. For a more efficient application of the method of radical background suppression (the method of active correlations), it is important to study the spectra of the recorded energy of implanted recoil nuclei with the possibility of modeling and parameterization of such spectra. This is what will make it possible in future experiments to choose the actual intervals of the corresponding energies when registering sequences of the ER-α type in real time.

To improve the method of radical background suppression (the method of active correlations), it is important to study the spectra of recorded recoil nuclei energies, and model and parameterize these spectra. This will allow for the selection of actual energy intervals during future experiments for registration of real-time ER-α type sequences.